\input harvmac
\input epsf


\def\p{\partial}
\def\em6{e^{-6\alpha t}}

\Title{}{\vbox{\centerline{Some Low Dimensional Evidence for the
Weak}
\medskip
\centerline{ Gravity Conjecture }}} \centerline{Miao
Li$^{1,2}$, Wei Song$^1$ and Tower Wang$^1$}
\medskip
\centerline{\it $^1$ Institute of Theoretical Physics}
\centerline{\it Academia Sinica, P. O. Box 2735} \centerline{\it
Beijing 100080} \centerline{\it and}
\medskip
\centerline{\it $^2$ Interdisciplinary Center of Theoretical
Studies} \centerline{\it Academia Sinica, Beijing 100080, China}

\medskip

\centerline{\tt mli@itp.ac.cn, wsong@itp.ac.cn}
\centerline{\tt wangtao218@itp.ac.cn}
\medskip

We discuss a few examples in 2+1 demensions and 1+1 dimensions
supporting a recent conjecture concerning the relation between the
Planck scale and the coupling strength of a non-gravitional
interaction, unlike those examples in 3+1 dimensions, we do not
have to resort to exotic physics such as small black holes.
However, the result concerning these low dimensional examples is a
direct consequence of the 3+1 dimensional conjecture.

\Date{Jan. 2006}

\nref\alnv{N. Arkani-Hamed, L. Motl, A. Nicolis and C. Vafa, ``The String Landscape,
Black Holes and Gravity as the Weakest Force," hep-th/0601001.}
\nref\cv{C. Vafa, ``The String Landscape and the Swampland," hep-th/0509212.}
\nref\md{M. Douglas and Z. Lu, ``Finiteness of volume of moduli spaces," hep-th/0509224.}
\nref\gth{G. 't Hooft, ``Under the Spell of the Gauge Principle," World Scientific.}
\nref\djt{S. Deser, R. Jackiw and G. 't Hooft, ``Three-dimensional Einstein Gravity:
Dynamics of Flat Space," Annals Phys.152 (19840 220.}
\nref\ap{A. M. Polyakov, ``Quark Confinement and Topology of Gauge Theories,"
Nucl. Phys. B120 (1977) 429.}
\nref\hs{J. Harvey and A. Strominger, ``Quantum Aspects of Black Holes," hep-th/9209055.}
\nref\kms{ S. Kachru, J. McGreevy and P. Svrcek, ``Bounds on masses of bulk fields
in string compactifications," hep-th/0601111.}

Given the confusing situation in constructing string solutions using the effective field theory
approach, it becomes important to derive constraints on the structure of viable effective field
theories from string theory. A remarkable conjecture was recently forward in \alnv\ stating that
for a $U(1)$ interaction with strength $g$, there must be a cut-off parametrically
smaller than $gM_{pl}$, where $M_{pl}$ is the 4 dimensional Planck mass. For a very small
coupling $g$, this can be a rather low energy scale
otherwise unconstrained in an effective field theory. Moreover, there must exist a charged
particle whose mass is smaller than $gM_{pl}$.
This line of approach starts with \refs{\cv,\md}, where the authors argue that the number of massless
fields must be finite.

The main argument in \alnv\ for the relation $m\le gM_{pl}$ valid for a light charged
particle relies heavily on doing away with
the problem of Planck scale remnants, this problem leads to global symmetries which are
supposed nonexistent in a theory of quantum gravity. Even we accept the statement that there
can not be too many Planck scale remnants as a proved one, the link presented in \alnv\ between
this statement and relation $m\le gM_{pl}$ is not rock solid, thus it is desirable to find
more evidence for this relation without resorting to exotic physics such as the remnant problem.

Rewriting $m\le gM_{pl}$ as $m\le g/\sqrt{G}$, the latter is a universal statement in any
dimensions. For instance, suppose we compactify the 4D theory in question to 3D or 2D on a
flat torus, if the original 4D theory is a consistent theory, we see no reason for the resulting low
dimensional theory not to be consistent. Now, both $\sqrt{G}$ and $g$ are reduced by a factor
$\sqrt{V}$, where $V$ is the volume of the torus, thus $m\le g/\sqrt{G}$ is still valid in
the low dimensional theory, with $g$ and $G$ get interpreted as the low dimensional gauge
coupling and the Newton constant. For instance, in 3D, $M_{Pl}=1/\sqrt{G}$ and we have
$m\le g\sqrt{M_{pl}}$, $g^2$ has the dimension of mass in 3D.

In this note we shall start with a couple of 3D examples and end with a 2D example.

The first example we consider is the Nielsen-Olsen vortex, it is a solution of the system
of a $U(1)$ gauge field coupled to a complex scalar with an action
\eqn\vac{S=\int d^3x\left( -{1\over 4}F_{\mu\nu}F^{\mu\nu}-D_\mu\bar{\phi}
D_\mu\phi-{\lambda\over 2}(\bar{\phi}\phi -F^2)^2\right),}
where the scalar has a charge $g$. The mass of a static vortex solution is \gth
\eqn\vm{m={2\pi m_W^2\over g^2}C_1(\beta),}
where $m_W^2=2g^2F^2$ is the W boson mass, $C_1(\beta)$ is a function of the dimensionless
ratio $\beta=\lambda/g^2$ and parametrically is of order 1.

We now couple the above system to the 2+1 dimensional gravity. We shall not try to find
out the exact solution of a vortex with the presence of gravity. For us, it is sufficient
to know that a mass of particle will generate a metric with a deficit angle. For a point-like
mass, the metric reads \djt
\eqn\ppm{ds^2=-dt^2+r^{-8Gm}[dr^2+r^2d\theta^2],}
where $m$ is the mass of the point particle, this metric is actually flat. The deficit angle
is $8\pi Gm$, if it exceeds $2\pi$, the location of the particle $r=0$ blows up to a circle
of infinite radius. Thus, we require $8\pi Gm< 2\pi$ or $m<1/(4G)$. Although we do not know
the exact metric generated by the vortex \vm, we expect that asymptotically there will be
a deficit angle $8\pi Gm$. Applying the mass formula \vm\ to the above inequality we derive,
parametrically
\eqn\tdie{m_W<g/\sqrt{G}.}
Thus, we can state that there is a charged particle (W boson) whose mass is bounded by
$g/\sqrt{G}$. If we go a step further to take $m_W$ as the cut-off of the the effective
$U(1)$ theory, we have $\Lambda<g/\sqrt{G}$. Of course the real cut-off can be different from
$m_W$, since we can imagine that the effective $U(1)$ theory is a descendant of a nonabelian
gauge theory with a spontaneous symmetry breaking scale different from $m_W$.

We could also consider other types of vortex solutions, the important point is that the
mass of a solution always scales as $m_W^2/g^2$, and our argument goes through.

Next, we consider a $U(1)$  theory descending from a $SU(2)$ theory.
The case of a monopole is discussed in \alnv. The mass of a monopole is $m=\Lambda/g^2$, the cut-off
$\Lambda\sim m_W$, here $m_W$ is the W boson mass in the 4D theory. The monopole has a
field-theoretic size $1/\Lambda$, its gravitational size is $Gm$, demanding the latter be smaller
than the former one deduces $\Lambda<g/\sqrt{G}$. The authors of \alnv\ argue that a monopole
should not become a black hole thus its gravitational size should be smaller than its
field-theoretic size. In 3D, a monopole becomes an instanton \ap\ if we take the Euclidean
time to be one of the original three spatial dimensions, and mass becomes action.
The action of an instanton is $m_W/g^2$, a dimensionless quantity since $g^2$ has a mass
dimension in 3D. In 3 dimensional Euclidean spacetime, there is no horizon. The inequality
$m_W<g/\sqrt{G}$ in 3D can be derived by the requirement that the gravitational size be smaller
than the field-theoretic size too. To see this, note that due to the Einstein equations
$G_{\mu\nu}=8\pi GT_{\mu\nu}$, there is a relation $R\sim GT$, $T$ is the trace of $T_{\mu\nu}$,
thus the Einstein action $1/(16\pi G)\int \sqrt{g}R\sim \int \sqrt{g}T$. This implies
that the gravitational action is the same order of the field theory action. Let the gravitional
size be $l_G$, the gravitational action is of order $l_G/G$, we therefore have
$l_G/G\sim m_W/g^2$ or $l_G\sim Gm_W/g^2$. Now, $m_W<g/\sqrt{G}$ follows from $l_G<1/m_W$.
Why should we demand the gravitational size of an instanton be smaller than its field-theoretic size? Apparently,
if $l_G>1/m_W$, the field theory can be no longer trusted in the neighborhood $r<l_G$, since
the gravitational field scales as $l_G/r$ and becomes strong in this neighborhood, however,
the monopole solution  locates well inside this neighborhood if $1/m_W<l_G$, we expect large
gravitational correction to the solution, and the original solution and its action can no
longer be trusted. Although this example is not as clean as the vortex case, we believe
that the above argument can be cast into a rigorous statement that the gravitational back-reaction
will destroy the monopole solution. Since there is no horizon involved in the Euclidean
solution, this is perhaps a better argument than the one in the 4D theory.

Our final example is a 2D solution. Ideally, it would be nice to generalize the kink solution
of a real scalar field to one coupled to the 2D dilaton gravity, it turned out the coupled system
is sufficient complex so no exact solution has been found. We resolve to consider a toy system
in which a $U(1)$ dipole is coupled to the 2D dilaton gravity. The action of the 2D dilaton
gravity coupled to a $U(1)$ gauge field reads
\eqn\tddg{S=\int d^2x\sqrt{-g}e^{-2\phi}\left(R+4(\p\phi)^2-\half F^2\right),}
without the $U(1)$ field, the system is amply discussed in for instance \hs.
We would like to find the solution with a pair of charges separated by a distance $l$.
In the flat spacetime, the solution is $F_{tx}=E=g(\theta (x)-\theta (x-l))$, namely,
$E=0$ outside $(0,l)$ and $E=g$ inside $(0,l)$. The energy of the dipole is $g^2l$.
Since the whole action \tddg\ is weighted by $\exp(-2\phi)$, after proper scaling, the
weak gravity statement is $g^2l<g$, or $gl<1$.

To find the exact solution of the dipole coupled to the dilaton gravity, choose the conformal
gauge in which the only non-vanishing metric component is $g_{+-}=-\half e^{2\rho}$, where we
used the light-cone coordinates $x^\pm=t\pm x$.
The equations of motion derived from action \tddg\ are
\eqn\tddge{\eqalign{&2\p_+\phi\p_-\phi -\p_+\p_-\phi-\half e^{-2\rho}E^2=0,\cr
&\p_+\p_-\rho-2\p_+\p_-\phi+2\p_+\phi\p_-\phi+\half e^{-2\rho}E^2=0,}}
supplemented with constraints derived from the e.o.m. of components $g_{++}$ and
$g_{--}$:
\eqn\tdct{\eqalign{&2\p_+\rho\p_+\phi -\p_+^2\phi=0,\cr
&2\p_-\rho\p_-\phi -\p_-^2\phi=0.}}
We expect a static solution. Outside $(0,l)$, the static solution, up to some overall
constants, is
\eqn\sts{e^{2\rho}=e^{2\phi}={1\over 1+cx}.}
We can choose $c$ properly on each side of the dipole to make the solution nonsingular
everywhere, since the scalar curvature $R=8e^{-2\rho}\p_+\p_-\rho\sim e^{2\rho}$ is
nonsingular. However, the coupling constant $e^\phi$ drops to zero at the infinity if
$c\ne 0$. To have a vacuum solution with a finite coupling at the infinity, we choose
$c=0$ outside $(0,l)$.

To find the solution inside $(0,l)$, we solve the constraints \tdct\ first, we have
$\phi'\sim e^{2\rho}$. The e.o.m. for $E$ is $\p_x[\exp(-2\phi-2\rho)E]=0$, thus
inside the dipole, $E=g\exp(2\phi+2\rho)$. Plug this and $\phi'\sim e^{2\rho}$ into
\tddge\ we find $2\phi=\rho$ and
\eqn\msl{e^{2\rho}={1\over 1-4gx},}
where we assumed that the $\rho=0$ to the left of the dipole. We can always demand
$\rho$ and $\phi$ be continuous at $x=0,l$, but the derivatives of these quantities
are not continuous. In other words, in order to have a honest solution, we need to
add sources to equations in \tddge\ and \tdct. Since inside $(0,l)$, $2\phi=\rho$,
no source is needed for the second equation in \tddge, this equation corresponds to
the e.o.m. of the dilaton. One can not have sources for the metric from the charges
since they couple only to the gauge field.
One can verify that it is sufficient to add terms
\eqn\sct{\int\sqrt{-g_{tt}}dtdx(g\delta(x)-g\delta(x-l))}
to the action in order to generate jumps in $\rho$ and $\phi$. The physical meaning
of these terms are just a pair of negative mass and position mass. It is not surprising
that these terms are needed: two opposite charges are attractive so we need to add
mass to generate repulsive gravitational force to have a balanced system. Finally,
from the solution \msl\ we deduce that $4gl<1$ in order to have a regular solution
inside the pair. This is our desired result.

We need to stress that the dipole system is not realistic. To have a similar realistic
system, we may consider a segment of open string, but it is a much more difficult
system to deal with since we perhaps need to quantize the string first. The dipole
size $l$ of the open string will be determined by the string tension. Nevertheless,
we expect that the result we obtained is still valid for this more realistic system.

As we mentioned earlier, it may be interesting to study the kink solution coupled
to the 2D dilaton gravity. Let $\lambda$ be the coupling of the quartic term of the
scalar field, $\mu$ the Higgs mass, then the mass of a kink is $\mu^3/\lambda$. Note
that $\lambda$ in 2D has a dimension of mass square. We conjecture that when $\mu>
\sqrt{\lambda}$ parametrically (we assume that  $\exp(-2\phi)$ is the
overall factor in the action), there is no regular solution in the coupled system.

To conclude, we have offered a few low dimensional examples supporting the weak gravity
conjecture of \alnv, and these examples surprisingly involve only classical gravity.
This fact may be related to the simplicity of quantum gravity in low dimensions, that is,
the UV effects are less important than the IR effects.

Note added: A new conjecture is made recently in a new paper \kms.

\bigskip

\noindent Acknowledgments

This work was supported by grants from CNSF.


\listrefs
\end